# Structural, Optical and Single-domain Magnetic Features of the Noncollinear Ferrimagnetic Nano-spinel Chromites $A\mathrm{Cr_2O_4}$ ($A$ = Ni, Co, and Mn)

Mohamed A. Kassem[1,2,3,*], Abdulaziz Abu El-Fadl[2,3], Ahmed M. Nashaat[1,2,3], Hiroyuki Nakamura[1],

1. Department of Materials Science and Engineering, Kyoto University, Kyoto 606-8501, Japan

2. *Department of Physics, Faculty of Science, Assiut University, 71516 Assiut, Egypt*

3. *Lab. of Smart Materials for Energy Futures, Faculty of Science, Assiut University, Assiut 71516, Egypt*

* Electronic email: makassem@aun.edu.eg

## Abstract

Spinel chromites $A\mathrm{Cr_2O_4}$ with inherent magnetic geometrical frustration usually exhibit a noncollinear ferrimagnetic ground state when $A^{2+}$ are magnetic ions, with possibly crystallite-size dependent intriguing magnetic features. Here, we report single-domain magnetic properties of $A\mathrm{Cr_2O_4}$ ($A$ = Ni, Co, and Mn) nanocrystals, with an average crystallite size of 18, 15 and 10 nm, exhibiting an optical energy gap of ~ 2.87, 3.05 and 2.9 eV, respectively. The temperature dependence of magnetization indicates the main bulk magnetic transitions with a commonly coexisting spin-glass-like state and finite-size effects on the noncolinear ferrimagnetic transitions. An anomaly observed at $T_s$ = 15, 24 and 10 K is attributed to the bulk magnetic transition to a canted antiferromagnetic state in $\mathrm{NiCr_2O_4}$ and incommensurate spiral orders in $\mathrm{CoCr_2O_4}$ and $\mathrm{MnCr_2O_4}$ NCs, respectively. A further bulk magnetic transition to a commensurate spiral order is observed for $\mathrm{CoCr_2O_4}$ NCs at a 'lock-in' temperature $T_l$ = 5 K much lower than that reported using bulk samples, while it is completely suppressed in the $\mathrm{MnCr_2O_4}$ NCs. Finite-size effects and single-domain magnetic behaviors indicated by anomalous temperature-dependences of the coercive field and the hysteresis-loop squareness, mainly driven by a magnetocrystalline anisotropy, are discussed in comparison to results reported using bulk counterparts.

## 1. Introduction

Insulating magnets of noncollinear spin order are platforms of related phenomena, such as multiferroicity, with possible functionalities in spintronic devices for high-density data storage [1]–[9]. Among the noncollinear magnetic insulators, ferrimagnetic (FIM) spinel chromites $ACr_2O_4$, are of great interests due to their exotic electronic and magnetic properties such as the large magnetooptical and magnetodielectric couplings with a spontaneous dielectric polarization [6], [7], [10], [11]. The $ACr_2O_4$ compounds crystalize in the normal spinel structure with the divalent $A^{2+}$ cations rest in the tetrahedrally coordinated A site while the $Cr^{3+}$ ions fully occupy the octahedrally coordinated B site forming a 3D pyrochlore sublattice, as schematically represented in Figs. 1(a), (b) for the common spinel $Fd\bar{3}m$ cubic structure. The spinel structure is generally known with its essential antiferromagnetic (AFM) exchange interactions $J_{A\text{-}A}$, $J_{B\text{-}B}$ and $J_{A\text{-}B}$ those cannot simultaneously be negative [12]. In the normal spinel chromites, the AFM $J_{Cr\text{-}Cr}$ lonely occurs with highly effective geometrical spin frustration when the $A^{2+}$ cations are nonmagnetic. In case the A site is occupied by magnetic ions, the negative $J_{A\text{-}Cr}$ becomes dominant and both $J_{A\text{-}A}$ and $J_{Cr\text{-}Cr}$ become positive resulting in a FIM order, however, the effects of the inherent geometrical spin frustration remain in FIM spinel chromites. The geometrical magnetic frustration is assumed as the key force for the stabilization of noncollinear spin order in these FIM insulating spinels [13]–[15].

The FIM spinel chromites $ACr_2O_4$ ($A$ = Ni, Cu, Fe, Mn, and Co) were extensively studied by magnetization, dielectric polarization, and microscopic measurements[6], [7], [10], [14], [16]. FIM chromites with Jahn-teller active $A^{2+}$ ions, such as $Ni^{2+}$, $Cu^{2+}$and $Fe^{2+}$, exhibit symmetry lowering from the cubic $Fd\bar{3}m$ to a tetragonal $I4_1/amd$ structure by a Jahn−Teller distortion at different temperatures, $T_{JT}$. For example, it occurs below room temperature at $T_{JT}$ = 140 K in case of $FeCr_2O_4$ (lattice parameters ratio, $c/a$ < 1)[17] while above room temperature at $T_{JT}$ = 310 K for $NiCr_2O_4$ ($c/a$ > 1) [18] and at $T_{JT}$ = 850 K for $CuCr_2O_4$ ($c/a$ < 1)[18]. At $T_{JT}$, the crystal-field splitting of the 3d levels are modified from those in the spinel cubic phase, as shown for $Ni^{2+}$ in Fig. 1(c). On the other hand, these FIM spinel chromites exhibit noncollinear FIM ground states, canted AFM states in $NiCr_2O_4$ and $CuCr_2O_4$ and conical FIM states in Fe, Co, and Mn spinel chromites. A FIM transition occurs at a Curie temperature $T_C$ ~ 74, 152, 97, 94 and 48 K for $A$ = Ni, Cu, Fe, Co, and Mn, respectively[5], [9], [19]. Except for $CuCr_2O_4$ that is a noncollinear FIM below $T_C$ [19], these FIM chromites exhibit a further magnetic transition from a collinear to a non-collinear FIM state below a temperature $T_s$ (< $T_C$ )[5], [14], [16]. In case of $A$ = Mn, Fe and Co, a spiral component emerges below $T_s$ ~ 19.5, 35, and 24 K, respectively, that changes from incommensurate to commensurate spiral spin structure in Co and Mn chromites at a ‘lock-in’ temperature, $T_l$ ~ 13 and 17.5 K, respectively[5], [7], [14], [16]. While in $NiCr_2O_4$, a collinear transverse antiferromagnetic (AFM) component develops, resulting in a canted AFM state, below $T_s$ ~ 31 K[16]. The successive structural and magnetic transitions between multiple states as described above are associated with exotic magnetic, optical and magnetoelectrical properties of these compounds. For instance, the magnetic transitions at $T_s$ are accompanied by an emergent polarization state, i.e., multiferroic effect, below $T_s$ in the FIM spinel chromites [1], [3], [6], [20]. In a sensitivity to the nearest-neighbor arrangement and the antiferromagnetic exchanges between *A*–Cr and Cr–Cr sites, changing the *A* cation type significantly changes the physical properties of the spinel chromites.

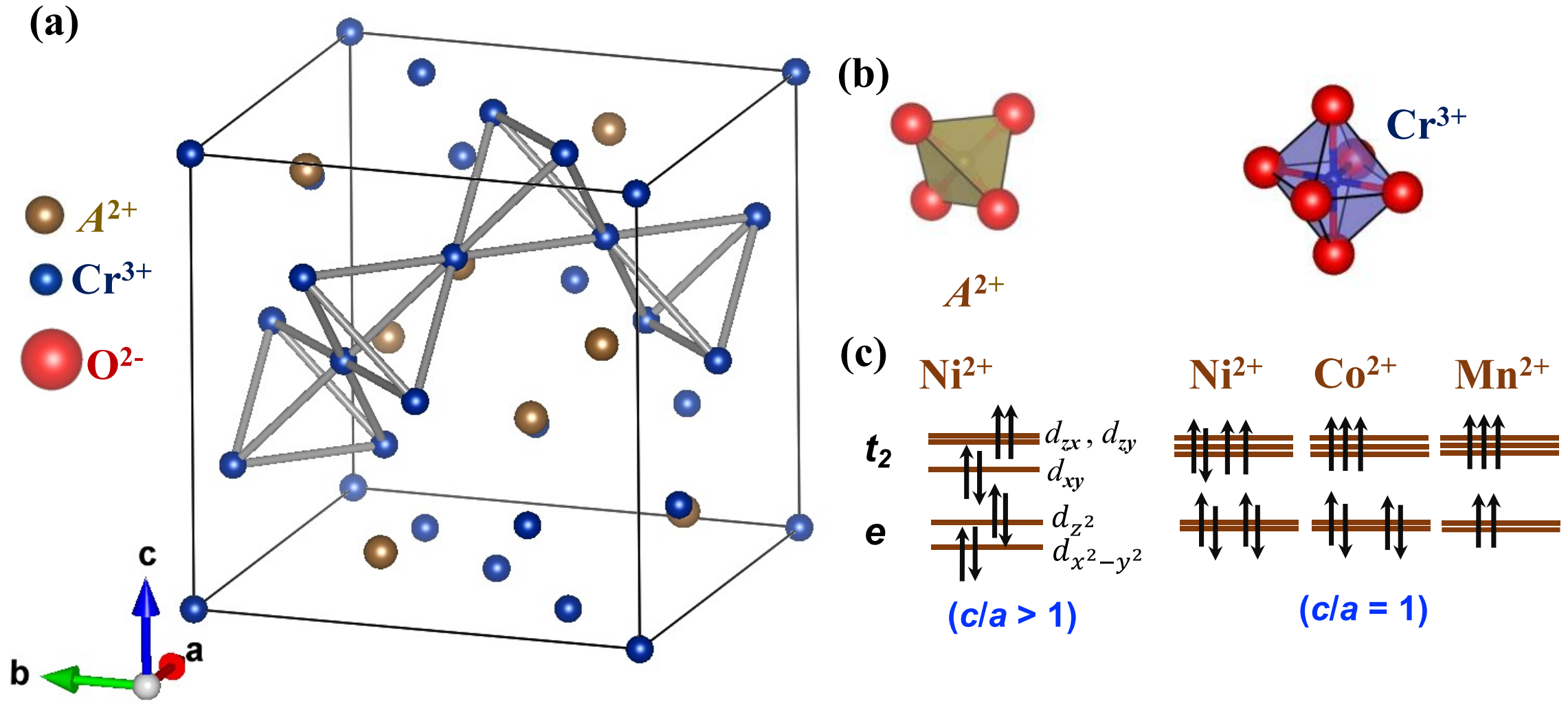

**Figure 1**. (a) Cubic crystal structure of the spinel chromites $ACr_2O_4$ schematically represented by a unit cell of the cationic ions $A^{2+}$and $Cr^{3+}$ with the $Cr^{3+}$ ions forming a pyrochlore sublattice. (b) Crystallographic polyhedrons of cations in the A and B sites, and (c) crystal-field splitting schemes of the 3d-levels and their occupation for $Ni^{2+}$, $Co^{2+}$, and $Mn^{2+}$ in the tetrahedral A site in the possible tetragonal ($c/a > 1$) and cubic ($c/a = 1$) spinel phases [21].

The magnetic properties of spinel chromites has been mainly studied using poly- and single-crystalline samples, however, the study of these materials as magnetic nanoparticles, hitherto rare relative to ferrites, can elucidate their property variations by finite size effects. Recently we have explored the magnetic properties of nanoparticles of the AFM spinel chromites $MCr_2O_4$ ($M$ = Zn, Mg and Cd) in comparison to the FIM $CoCr_2O_4$ nanoparticles [22]–[24]. In this manuscript, we study the optical and magnetic properties of single-phase and high-quality nanocrystals (NCs) of $ACr_2O_4$ ($A$ = Ni, Co, and Mn) with average particle (and crystallite) sizes of 15, 13 and 10 nm, respectively, synthesized by a facile microwave combustion method. The spectroscopic characteristics, corresponding to the ionic bonds (vibration) and electronic bands (transition), are studied by FT-IR and UV-Vis absorbance spectroscopies. Finite size effects on the bulk magnetic transitions and single-domain magnetic properties including high and largely temperature-dependent coercivity are explored. The observed single-domain nanomagnetism of these spinel noncollinear FIM chromites are discussed with an interrelation between their structural and optical characteristics, and in a comparison view to the results reported using bulk poly- and single crystalline samples.

## 2. Experimental procedures

### 2.1. Synthesis

Starting with salts of a divalent $A^{2+}$ cation, $Ni(NO_3)_2.6H_2O$, $Mn(NO_3)_2.6H_2O$, or $Co(NO_3)_2.6H_2O$, and trivalent $Cr^{3+}$ cation, $Cr(NO_3)_3.9H_2O$, nanoparticles of $ACr_2O_4$ ($A$ = Ni, Co or Mn) have been synthesized by a microwave combustion[24]. Glycine $NH_2CH_2COOH$ was employed in this synthesis as a combustion fuel that burns at a sufficient temperature initiating the reaction and completing the product oxidation process. As well as cations source, metal nitrates act as oxidant, while the fuel is a source of carbon and hydrogen and act as a partial reducing agent usually yielding gaseous products such as $CO_2$ and $H_2O$. However, the fuel ratio is crucial for the synthesis and is selected based on a required complete redox reaction[25]. The ratio of glycine to complete the reaction is calculated with respect to the gaseous product from the combustion[26], [27] and this reaction synthesis of $ACr_2O_4$ can be described as:

$$A(\mathrm{NO_3})_2.6\mathrm{H_2O} + 2\,\mathrm{Cr(NO_3)_3.9H_2O} + \delta\,\mathrm{C_2H_5O_2N} + \left(\frac{9\delta-40}{4}\right)\mathrm{O_2} \xrightarrow{\Delta} A\mathrm{Cr_2O_4} + \left(\frac{5\delta+48}{2}\right)\mathrm{H_2O}\uparrow + \left(\frac{\delta+5}{2}\right)\mathrm{N_2}\uparrow + 2\delta\,\mathrm{CO_2}\uparrow, \tag{1}$$

where $A$ is a divalent cation and $\delta$ is the amount required of the fuel glycine. The divalent and trivalent metal nitrates were mixed with glycine, $NH_2CH_2COOH$, in a molar ratio of 1: 2: $\delta$, respectively. Based on a possible reaction described by equation (1), oxygen appears as a reactant (from air) or byproduct gas when the glycine ratio is different from $\delta$ = 4.44. In our synthesis, glycine molar ratio of about $\delta$ = 2.4 was added in the synthesis of $NiCr_2O_4$ and $CoCr_2O_4$ while it was almost doubled while the synthesis of $MnCr_2O_4$ in a total mass of reactants of about 6 g for each pot synthesis. The mixed raw materials were dissolved in double distilled water by means of magnetic stirrer at room temperature till we obtained homogenous clear solution. The mixture solution was then introduced to a microwave oven operating for ~15 mins at a power of 800 watt. After complete evaporation of water, combustion occurs resulting in a dry fluffy solid product that was ground and prepared for characterization and measurements without further heat treatments.

### 2.2. Characterization and measurements

The structure at room temperature has been investigated by powder x-ray diffraction (XRD) using x-ray diffractometer (X'Pert Pro, PANalytical) with Cu $K_{\alpha 1}$ radiation ($\lambda$ = 1.5405 Å) monochromated by a Ge (111)-Johansson-type monochromator. The microstructure and the particles shape and size have been investigated by a JEOL JEM-2100F electron microscope. An attached unit of energy dispersive x-ray spectroscopy (EDX-TEM) has been employed to investigate the elemental compositions in many particles of each sample. FTIR spectroscopy has been performed on the samples via a NICOLET FTIR 6700 spectrometer for a pellet of each sample mixed with KBr powder. The measurements were performed in the wavenumber range 400-800 $cm^{-1}$. The absorption spectra of the samples were collected by UV-Vis spectrophotometer (Jasco v-750) in the wavelength range 200-800 nm for particles suspended in Dimethyl sulfoxide (DMSO) solvent with equivalent concentrations. For this purpose, equal amounts of the suspended nanoparticles were treated by ultrasonic waves directly before measurement for absorbance in a homogeneous suspension of the particles.

Magnetization has been measured by using a superconducting quantum interference device (SQUID) MPMS5 (Quantum Design) in the temperature range of 2 – 300 K with magnetic fields up to 5 T for each sample free particles charged in a gelatin capsule. Temperature dependencies of zero-field-cooled (ZFC) and field-cooled (FC) magnetization were measured at 0.01 and 1 T after waiting for thermal equilibrium at each temperature. The temperature dependencies of the spontaneous magnetization, coercivity, and the observed magnetic squareness ratio are in-correlation discussed for these chromite nanocrystals with finite size effects in comparison to reported results using samples in bulk forms.

## 3. Experimental Results

### 3.1. Structure and particle size

The powder XRD patterns of the synthesized nanoparticles were refined by means of RIETAN-FP software [28]. The XRD patterns observed at room temperature and corresponding Rietveld refinement results are shown in Fig. 2(a-c). The refinement reveals the formation of a spinel-structured single phase. The structural analysis of the XRD patterns indicated the formation of a cubic spinel structure for the three compounds nanoparticles, with peaks indexed in Fig.2(a). Although the bulk spinel chromite $NiCr_2O_4$ crystallize in a tetragonal structure at room temperature[29], we found $NiCr_2O_4$ fine NCs crystalize in a cubic structure at room temperature which agrees with previously reported results[30]. The refinement has been performed with assuming the cubic space group $Fd\bar{3}m$ (setting #2) with sites 8a (1/8, 1/8, 1/8) and 16d (½, ½, ½) for atoms at the tetrahedrally coordinated (A) and octahedrally coordinated (B) sites, respectively, while oxygen atoms occupy the 32e ($x$, $y$, $z$) site. Assuming stochiometric compositions, the lattice constant ($a$), oxygen position parameter ($u = x = y = z$), cations distribution among the A and B sites indicated by ions occupancy ($g$) and peaks shape parameters (FWHM, position, intensity, etc.) were refined. The tendency of occupying different sites by the cations $A^{2+}$ and $Cr^{3+}$ is dependent on the calculated stabilization energy; simply the highest magnitude of crystal field stabilization energy (CFSE) when the cations freely occupy possible sites [31]. We found that the refinement assuming free occupancy of both cations $A^{2+}$ and $Cr^{3+}$ in the A and B sites results in almost $g$ = 1 for both $A^{2+}$ in the A site, 8a (1/8, 1/8, 1/8), and $Cr^{3+}$ in the B site, 16d (½, ½, ½), in a normal spinel configuration for the three compounds, that agrees with the well-known high CFSE of $Cr^{3+}$ in the octahedral site [32], [33]. The main fitting parameters and refinement results are presented in Fig. 2(a-d) and Table 1. [34], [35] Another confirmation of the lack of cations inversion in the $ACr_2O_4$ ($A$ = Ni, Co, and Mn) nanoparticle spinel structure can be achieved by comparing the experimental value of lattice constant $a$ with that calculated for a normal spinel $ACr_2O_4$ as functions of the cation radius of $Ni^{2+}$, $Co^{2+}$ and $Mn^{2+}$ at the tetrahedral site. The theoretical lattice constant $a_{\mathrm{cal.}}$ is calculated using the equation [36]:

$$a_{\mathrm{cal.}} = \frac{8[(r_{\mathrm{A}}+r_{\mathrm{O}})+\sqrt{3}(r_{\mathrm{Cr}}+r_{\mathrm{O}})]}{3\sqrt{3}}, \tag{2}$$

where $r_{\mathrm{A}}$ is the radius of a cation in the tetrahedral A site, $r_{\mathrm{Cr}}$ is the radius of the $Cr^{3+}$ in the octahedral B site, and $r_{\mathrm{O}}$ is the radius of the oxygen anion, assumed as reported by Shannon[35]. Both experimentally observed and theoretically calculated lattice constants are presented in Fig. 2(d) and Table 1. The experimental lattice constant almost matches the theoretical values calculated by assuming the normal spinel structure and both show linear dependence on the ionic radius of $A^{2+}$ in a tetrahedrally coordinated site, as shown in Fig. 2(d).

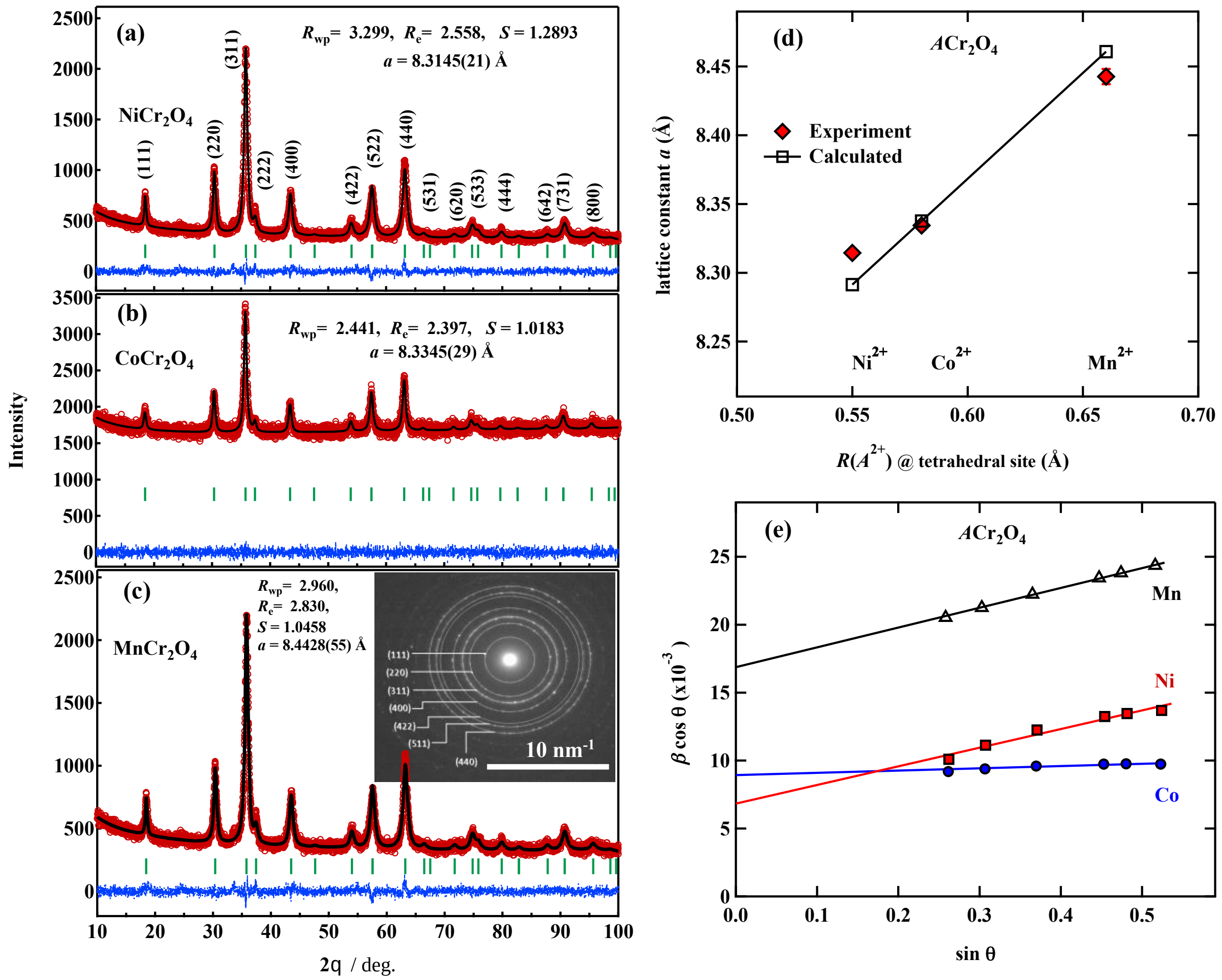

**Figure 2**. Room-temperature powder XRD patterns and Rietveld refinement for **(a)** $NiCr_2O_4$, **(b)** $CoCr_2O_4$ and **(c)** $MnCr_2O_4$ nanoparticles with labels for corresponding weighted-profile and expected refinement reliability indicators ($R_{wp}$ *and* $R_e$), goodness-of-fit factor $S = R_{wp}/R_e$ and lattice constant, $a$. Open circles represent the experimental data, while solid lines for calculated patterns, dashed lines for the difference between them, and vertical bars for corresponding Bragg diffraction angles. Inset of **(c)** shows a selected area electron diffraction (SAED) of $MnCr_2O_4$ NCs. **(d)** the variation of both experimentally measured and theoretically calculated lattice constant, $a$, with the ionic radius of $A^{2+}$ ions in the tetrahedral site, $R(A^{2+})$. **(e)** Williamson-Hall plots, and their linear fits, as estimated from the XRD diffraction patterns.

The powder XRD patterns of the different compounds show variations in the peaks' broadening ranging from broad peaks for $MnCr_2O_4$ to relatively sharp peaks for $NiCr_2O_4$, reflecting the larger crystallite size of $NiCr_2O_4$ compared to the other two samples. The crystallite size can be estimated from the powder XRD patterns, with correction for the contribution of lattice strain to the peak broadening, by employing the Williamson-Hall (W-H) method of a formula [34]:

$$\beta_{hkl} \cos\theta = \frac{k\lambda}{D_{hkl}} + 4\varepsilon \sin\theta \tag{3}$$

where $\beta_{\mathrm{hkl}}$ is the full width at half maximum of the ($hkl$) peak, $\theta$ is its Bragg's angle, $\lambda$ is the x-ray wavelength, $k$ is a constant (taken as 0.9), $\varepsilon$ is the strain in the lattice unit cell and $D_{hkl}$ is the corresponding crystallite size. The average crystallite size $D_{\mathrm{XRD}}$ can be calculated from a linear fitting of the W-H plot between $\beta_{\mathrm{hkl}} \cos\theta$ and $\sin\theta$ for different peaks in the XRD pattern. The W-H plot, shown in Fig. 2(e), elucidates linear fits for the three compounds data according to equation (3) and the estimated $D_{\mathrm{XRD}}$ values are presented in Table 1.

High resolution TEM images of the synthesized nanoparticles are presented in Fig. 3. The synthesized particles were confirmed to be within the nanoscale with different particle size distributions. The obtained largest particles of the three compounds are sized below 30 nm. As shown in Figs. 3, nanoparticles with particle size distribution ranges of 10 –30, 5 – 27, and 3 – 18 nm and average particle size $D_{\mathrm{TEM}}$ of ~16, 13 and 10 nm are observed for $NiCr_2O_4$, $CoCr_2O_4$, and $MnCr_2O_4$ NCs, respectively. The average particle size matches the crystallite size estimated from XRD analysis indicating successful growth of nanocrystals (NCs). As an example, high magnification images show the fringes of $NiCr_2O_4$ NCs indicating a high degree of crystallinity. The $MnCr_2O_4$ NCs are grown in lowest particle size in a relatively narrow size distribution, Fig. 3(c), which indicates that using a large amount of the fuel while the microwave-induced-combustion synthesis may reduce the particle size of the synthesized chromite nanoparticles.

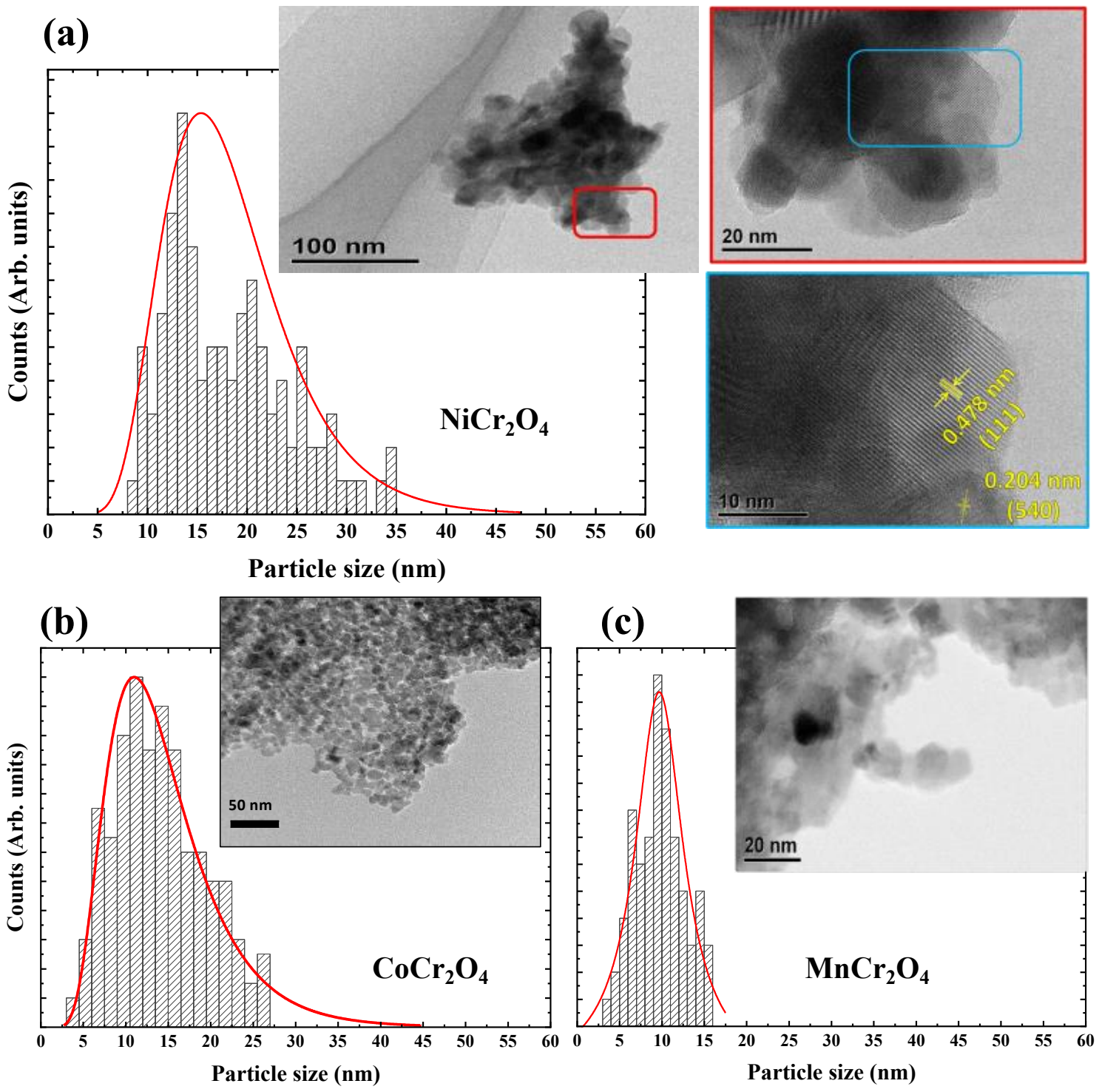

**Figure 3.** (HR)TEM micrographs for (a) $NiCr_2O_4$ at different magnifications showing the interplanar fringes corresponding indexed planes, and for (b) $CoCr_2O_4$ and (c) $MnCr_2O_4$ nanoparticles, all with particle size histograms.

**Table 1.** Experimental structural parameters: lattice constant ($a$), oxygen position parameter ($u$) crystallite size ($\boldsymbol{D}_{\mathbf{XRD}}$) and lattice strain ($\boldsymbol{\varepsilon}$), estimated based on XRD Rietveld refinement calculated lattice constant ($\boldsymbol{a}_{\mathbf{cal.}}$) of $ACr_2O_4$ ($A$ = Ni, Co, and Mn) NCs, and TEM-determined particle size ($\boldsymbol{D}_{\mathbf{TEM}}$).

| **Sample** | $a_{exp.}$ (Å) | $u$ | <$D_{XRD}$> (nm) | $\varepsilon \times 10^{-3}$ | $a_{cal.}$ (Å) | <$D_{TEM}$> (nm) |
|---|---|---|---|---|---|---|
| **$NiCr_2O_4$** | 8.314(2) | 0.2582(4) | 18.1(±1.3) | 2.99(3) | 8.291 | 16(±2) |
| **$CoCr_2O_4$** | 8.334(3) | 0.2624(6) | 15.5(±0.28) | 0.41(9) | 8.338 | 13(±1) |
| **$MnCr_2O_4$** | 8.434(65) | 0.2627(5) | 8.2(±0.05) | 3.64(6) | 8.461 | 10 (±1) |

### 3.2. Spectroscopic Characteristics: FT-IR and UV-Vis spectroscopies

Further identification evidence of the spinel phase can be provided by an investigation of the IR absorbance spectra corresponding to its cation-anion bonds vibrational modes. Mainly, there are two characteristic stretching vibrations in the spinel structure attributed to its cation-anion bonds in the available tetrahedrally and octahedrally coordinated sites. The main two bands occur above 400 $cm^{-1}$ are the stretching vibration frequency $\upsilon_1$ corresponding the *A*-O bond at the tetrahedral site and the frequency $\upsilon_2$ of the Cr-O bond at the octahedral site [37]. Other spinel characteristic bands can be detected in the far-IR region below 400 $cm^{-1}$ [38], however, the detection of $\upsilon_1$ and $\upsilon_2$ can be considered as strong evidence for the formation of the spinel phase. Furthermore, the shape of the developed vibrational bands may indicate some information about the cation distributions between the A and B sites.

Figure 4 shows the FT-IR spectra of $ACr_2O_4$ (*A* = Ni, Co, and Mn) NCs exhibiting the main characteristic bands around 600 $cm^{-1}$ and 500 $cm^{-1}$. The assignments of the observed vibrational bands are presented in Table 2. The relative position of the two characteristic bands slightly shifts to lower frequencies with varying the divalent cation from $Co^{2+}$, $Ni^{2+}$ to $Mn^{2+}$, respectively. Characteristic bands of bulk $CoCr_2O_4$ were observed at 630 and 525 $cm^{-1}$, for the tetrahedral and octahedral cation-anion bonds, respectively, those bands slightly shift to higher wavenumbers for $NiCr_2O_4$[39]. We found a decrease of $\upsilon_1$ and $\upsilon_2$ from 630 $cm^{-1}$ to 611 $cm^{-1}$ and from 525 $cm^{-1}$ to 491 $cm^{-1}$, in the case of $CoCr_2O_4$. This shift in bands position can be attributed to the smaller particle or grain size as previously reported[40]. Similar band shift for powdered samples of ionic crystals was explained by a depolarization field related to the small crystallites[41].

The vibrational characteristic bands in Fig 4 do not show any splitting, however, a small shoulder is observed around 675 $cm^{-1}$ and it is more prominent for the $NiCr_2O_4$ and $CoCr_2O_4$ NCs, that agree with previous observations[30], [42]. The formation of mixed cations in the octahedral site are excluded in chromites, as $Cr^{3+}$ cation have high CFSE in the octahedral site compared to other cations in different spinel compounds. A possible explanation is the scattering from randomly oriented crystallites with large distribution in shape and size in $NiCr_2O_4$ and $CoCr_2O_4$ as seen in the TEM results[43].

The force constant $K$ for the tetrahedral site ($K_{Tet}$) and octahedral site ($K_{Oct}$) can be determined from the relation[44]:

$$K = 4\pi^2 c^2 \upsilon^2 \mu. \quad (4)$$

Where $c$ is the speed of light, $\upsilon$ is the vibration band frequency, and $\mu$ is the reduced mass of the corresponding cation and anion with masses $m_C$ and $m_O$, respectively. which can be simply calculated as $\mu = \frac{m_O m_C}{m_O + m_C}$ [45]. In this normal spinel structure, the reduced mass at the octahedral site $\mu_{Oct}$g is common and equals to 2.032 $\times10^{-23}$ g, while $\mu_{Tet}$ values at the tetrahedral site are 2.087 $\times10^{-23}$ g, 2.057 $\times10^{-23}$, 2.089 $\times10^{-23}$ g for *A*= Ni, Mn, and Co, respectively. The obtained force constant values are listed in Table 2. We note that the vibrational energies and corresponding values of the force constant do not show systematic trends with the cationic radii of the divalent cations and their effect on the bond length. However, the difference $(\upsilon_1 - \upsilon_2)$ gets along with the radii of the cations used which is a further indication of the formation of almost complete normal spinel.

The optical absorption characteristics of the synthesized nanoparticles were investigated in the UV-Vis region of a wavelength range 300-800 nm. The absorbance (*A*) spectra are shown in the inset of Fig. 4(b). The absorption coefficient $(\alpha)$ was estimated using the Beer-Lambert law [46]:

$$A = \alpha\, c\, l, \tag{5}$$

and by considering the theoretical density $\rho$ in gm/cm$^3$ of the material, the absorption coefficient $(\alpha)$ in cm$^{-1}$ can be obtained as:

$$\alpha = \frac{A\,\rho}{c\,l}. \tag{6}$$

where *c* is the concentration of the sample in the solvent and *l* is the width of the suspension. The optical energy gap can be determined from the absorption coefficient $\alpha$ and the incident photon energy $h\upsilon$ using the Tauc's equation [47]:

$$\alpha h\upsilon = B(h\upsilon - E_g)^n. \tag{7}$$

Where *B* is a probability constant, and *n* is an exponent that determines the transition type. The optical performance of the samples demonstrates a typical allowed direct transition behavior with *n*=1/2.

Tauc's plots, $(\alpha h\upsilon)^2$ vs $h\upsilon$, are shown in the main panel of Fig. 4(b) for the three compounds NCs. The values of $E_g$ were estimated by linear extrapolation of data at high photon energy region, where the optical absorption is stable, to $(\alpha h\upsilon)^2$ = 0. The estimated values of $E_g$, presented in Table 2, are consistent with results of optical conductivity investigations for bulk chromites with a direct bandgap of $CoCr_2O_4$ nanoparticles reported as ~3.1 eV [5], [9], [48]. It is reported that a spinel chromite with a large optical energy gap undergoes first-order phase transition into spiral spin ordering at a low temperature[5], which is in consistence with the spiral spin transition observed in $CoCr_2O_4$ rather than in $MnCr_2O_4$ fine NCs as described below in section 3.3.

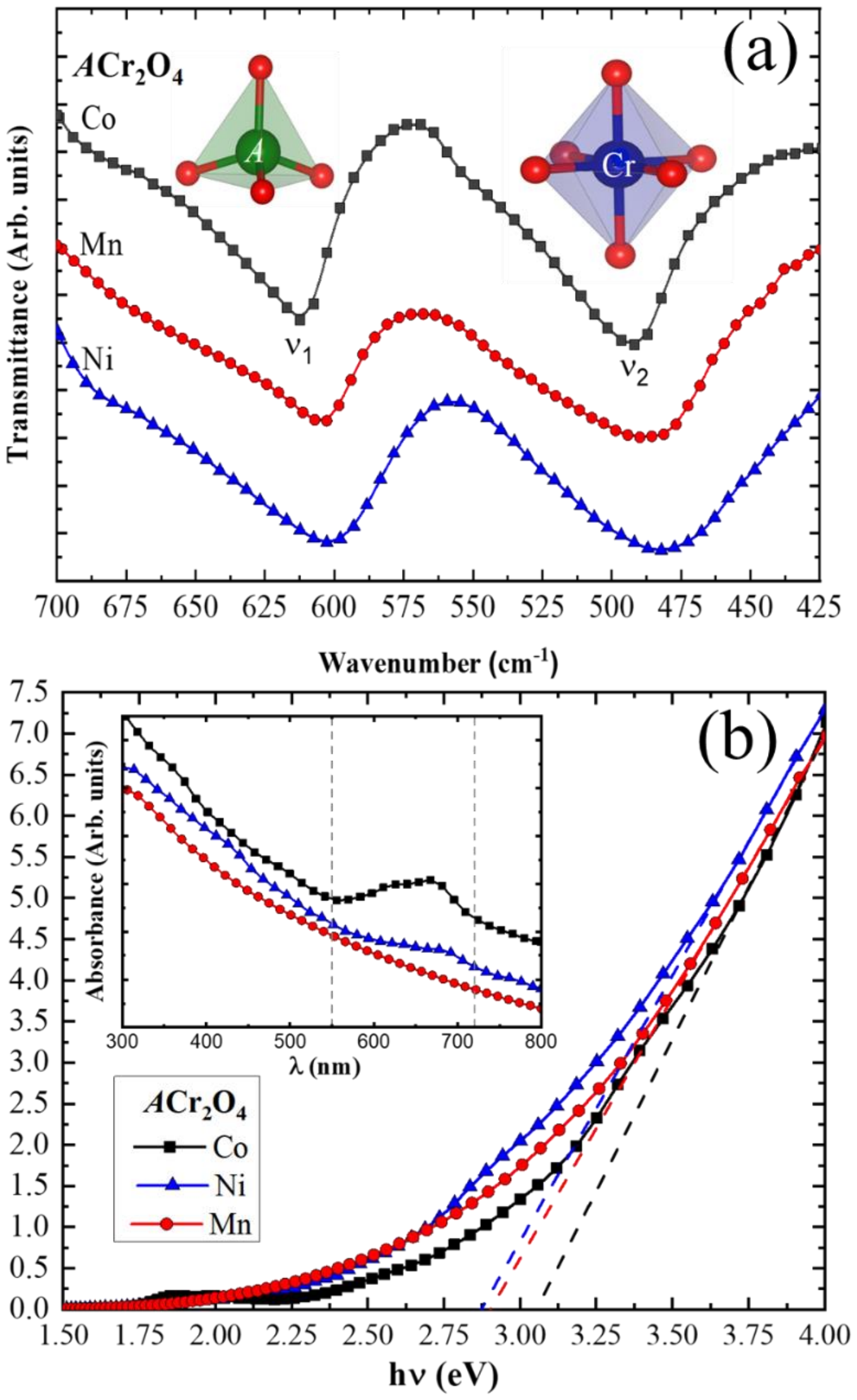

**Figure 4**. **(a)** FTIR spectra of $NiCr_2O_4$, $CoCr_2O_4$ and $MnCr_2O_4$ NCs presented in the wavenumber range between 700 nm and 400 $cm^{-1}$. **(b)** Tauc's plots established from UV-Vis absorbance data of $NiCr_2O_4$, $CoCr_2O_4$ and $MnCr_2O_4$ NCs, assuming allowed direct electronic transitions. Dashed lines represent extrapolations to corresponding optical energy gap values for each compound NCs. Inset shows the dimensionless absorbance spectra with small vertical offsets for clarity.

Additional absorption bands are observed at low photon energies that vary in position and intensity for the different three chromites, as seen in the inset of Fig.4(b). $CoCr_2O_4$ and $NiCr_2O_4$ exhibit absorption bands in the wavelength region 600 – 700 nm while $MnCr_2O_4$ nanoparticles do not develop anomaly. The absorption band of the $CoCr_2O_4$ is an asymmetric band with maxima around ~665 nm, more examination of this band draws the attention to a damped maxima around ~620 nm which overlapped and gave the asymmetric shape of the band. These are characteristic bands for the spin allowed transitions of $Co^{2+}$ in the spinel structure [42], [49] and can be assigned using Tanabe-Sugano diagrams of spinel crystal field[50]. The bands forming the big hump in the $CoCr_2O_4$ spectrum give evidence of the $d^7$ configuration of $Co^{2+}$, and they can be assigned to $^4A_2(F) \rightarrow {}^4T_1(P)$ transitions of the $Co^{2+}$ ions in tetrahedral coordination. The $^4A_{2g} \rightarrow {}^4T_{2g}$ and $^4A_{2g} \rightarrow {}^2T_{1g}$ transitions of $Cr^{3+}$ in octahedral coordination occur at approximate positions which give more probability of overlapping between the transitions in the spectrum, that may justify the broadness of the bands in the spectrum. In the case of $NiCr_2O_4$ weaker band is observed in the absorbance spectrum also broadened with maxima around ~685 nm, which can be attributed mainly to the 3d-3d transition between $Ni^{2+}$ tetrahedrally coordinated ions. The broadening of the band developed for $NiCr_2O_4$ spectrum may be attributed to the splitting in the 3d level of $Ni^{2+}$ cation as highlighted in Fig. 1(c), when compared to bulk the room temperature phase is tetragonal but for our samples the main phase is cubic which affects the symmetry and hence the optical transitions between bands in the UV-Vis spectral range. The splitting in the 3d energy levels in case of $NiCr_2O_4$ may justify the intensity of the optical band when compared to $CoCr_2O_4$ nanoparticles. In the case of $MnCr_2O_4$ sample there were no noticeable bands. Such behavior was predicted as the 3d-3d transitions are spin forbidden for the tetrahedrally coordinated $Mn^{2+}$[9].

**Table 2.** Spectroscopic characteristics of $ACr_2O_4$ ($A$ = Ni, Co, and Mn) NCs.

| Sample | FTIR Assignments | | Force constant | | | Optical Energy gap |
|---|---|---|---|---|---|---|
| | $\upsilon_1$ (cm$^{-1}$) | $\upsilon_2$ (cm$^{-1}$) | $K_{\text{Oct}}$ ($\times 10^5$dyne/cm) | $K_{\text{Tet}}$ ($\times 10^5$dyne/cm) | $\upsilon_1 - \upsilon_2$ (cm$^{-1}$) | $E_g$ (eV) |
| **$CoCr_2O_4$** | 611 | 487 | 1.729 | 2.753 | 120 | 3.055(13) |
| **$MnCr_2O_4$** | 604 | 491 | 1.701 | 2.649 | 117 | 2.903 (4) |
| **$NiCr_2O_4$** | 602 | 481 | 1.659 | 2.670 | 121 | 2.872 (6) |

### 3.3. Magnetic properties

To investigate the magnetic properties of $ACr_2O_4$ NCs, we have measured their magnetization as a function of temperature at low and high magnetic fields and its hysteresis loops at different temperatures after ZFC and FC processes. Figure 5(a-c) show the temperature dependence of the magnetic susceptibility, $M(T)/H$, measured at $H$ = 100 Oe after ZFC and FC processes. The magnetic transitions are determined from the temperature dependence of the FC magnetization derivative, d$M$(T)/d$T$, correspondingly shown in the insets of Figure 5(a-c). While the temperature decreases, the onset of FIM ordering is observed at a Curie temperature $T_C$ = 69, 92 and 40 K, as observed at low fields of 100 Oe, for $NiCr_2O_4$, $CoCr_2O_4$, and $MnCr_2O_4$ NCs, respectively. The observed $T_C$ values agree with those reported using bulk polycrystalline samples[6]. A bifurcation of the ZFC curve from the FC one is commonly observed below a freezing temperature $T_f$ of ~68, 88 and 20 K observed at 100 Oe, that decreases to 60, 75 and 7 K at 10 kOe, for $NiCr_2O_4$, $CoCr_2O_4$, and $MnCr_2O_4$ NCs, respectively. It is noticed that the spin freezing is almost released by applying a field of 1 T for the Mn chromite NCs, unlike the other two chromite NCs, as shown in Figs. 5(d)–(f). This may indicate a spin-glass-like (SGL) state rather than a surface disorder in Ni and Co chromite nanoparticles [14], [51], [52]. Based on magnetization and ac susceptibility analysis results, we have revealed the SGL in $CoCr_2O_4$ NCs[23]. The observed SGL state may result in large coercivity in these chromite NCs. With further decrease of temperature, a subsequent anomaly which can be attributed to a bulk noncollinear magnetic transition is observed at a temperature $T_s$ ~15 K for $NiCr_2O_4$, 24 K for $CoCr_2O_4$ and is barely observed around 10 K for $MnCr_2O_4$ NCs.

The transition of $NiCr_2O_4$ at $T_s$ is attributed to a bulk canted antiferromagnetic ordering by an emergent transverse AFM component [16]. We observed that $T_s$ around 15 K for $NiCr_2O_4$ NCS of average size ~15 nm is significantly lower than that reported for particles of size larger than ~30 nm as well as bulk samples with $T_s$ of about 30 K [6], [14], [53]. Additional slight anomalies observed at temperatures up to $T_s^*$ ~30 K, seen in $M(T)$ measured at 100 Oe as in the inset of Fig. 5(a), can be ascribed to a canted-AFM transition in particles with larger size, as elucidated by a relatively wide particle-size distribution (8 - 30 nm) shown in Fig. 3(a). On the other hand, the anomaly at $T_s$ ~24 K for $CoCr_2O_4$ NCs is attributed to a transition from the FIM state to an incommensurate FIM spiral order [14], [54]. With further decrease of the temperature, another anomaly is observed for $CoCr_2O_4$ NCs at $T_l$ ~ 5 K which is corresponding to a 'lock-in' transition to a commensurate FIM spiral order at which the spiral order becomes fully developed [14], [54]. Although the incommensurate spiral transition temperature $T_s$ of $CoCr_2O_4$ NCs is almost identical to that of its bulk forms, the 'lock-in' temperature $T_l$ becomes much lower than corresponding bulk values, probably, by structural and finite-size effects. Similar results were observed for $CoCr_2O_4$ NCs with a critical size limit of ~5 nm for the spin-spiral order and disappearance of the 'lock-in' transition for sizes up to ~14 nm[55]. For the other spiral FIM chromite $MnCr_2O_4$ NCs, with average crystallite size of ~10 nm, the anomaly at $T_s$ is barely seen around 10 K only at low fields of 100 Oe, while the 'lock-in' transition completely disappears. These results indicate that our FIM spinel chromite NCs are very close to the crystallite-size limit corresponding a suppressed noncollinear FIM order. Figs. 5(d)–(f) show the temperature dependence of magnetization data measured at a higher magnetic field of 10 kOe. We observe that $T_s$ is completely suppressed in the Mn chromite NCs while it is robust against high magnetic fields in Co and Ni chromites as the case of bulk samples[15], [56], [57].

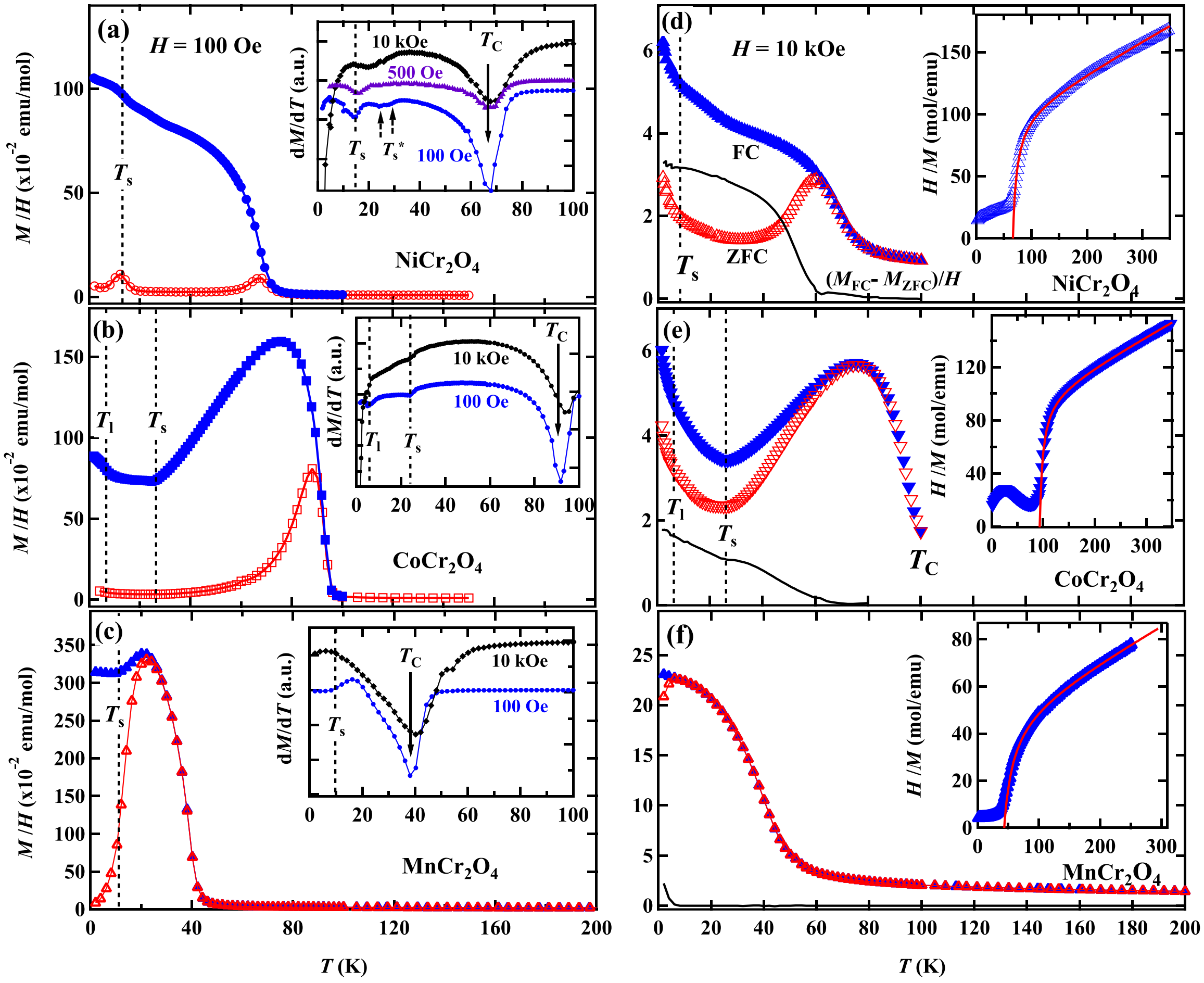

**Figure 5.** Temperature dependence of the magnetic susceptibility ($M/H$) measured at 0.1 and 10 kOe after ZFC (open) and FC (solid) processes for (a), (d) $NiCr_2O_4$; (b), (e) $CoCr_2O_4$ and (c), (f) $MnCr_2O_4$ NCs, respectively. Corresponding insets of (a) – (c) show the magnetic transition temperatures appear in the derivative of the FC magnetization, d$M$/d$T$($T$), at different fields, while insets in (d) – (f) are the $T$-dependence of the inverse susceptibility ($H/M$) measured at 10 kOe after a FC process. Solid lines in (d) – (f) are the difference between the FC and ZFC magnetizations measured at 10 kOe, while in (d) – (f) insets are the best fits to Eq. (8), in the text.

The insets of Figs. 5(d)–(f) show the inverse susceptibility, $\chi^{-1} = H/M$, as a function of temperature for the three compounds NCs, respectively. In the Néel molecular-field theory, the magnetization of FIM materials of nonequivalent sublattices behaves above $T_C$ with a hyperbolic characteristic of the inverse susceptibility, $\chi^{-1} = H/M$, as [58], [59]:

$$\chi^{-1} = \frac{T-\theta_W}{C} - \frac{\xi}{T-\theta'}\,. \tag{8}$$

The first term in Eq. (8) is the hyperbolic high-temperature linear part described in a Curie-Weiss (CW) form, with $\theta_W$ is the Weiss temperature and $C$ is the Curie constant The second term is the

**Table (3):** Particle size and magnetic parameters and transitions of $ACr_2O_4$ ($A$= Ni, Co, Mn) nanocrystals compared to previously reported data of bulk poly- and single-crystalline samples.

| **Chromite** | **$NiCr_2O_4$** | | **$CoCr_2O_4$** | | **$MnCr_2O_4$** | |
|---|---|---|---|---|---|---|
| **Form** | **NCs [this work]** | **Bulk [a] [refs.]** | **NCs [this work]** | **Bulk [a] [refs.]** | **NCs [this work]** | **Bulk [a] [refs.]** |
| **$D_{TEM}$ (nm)** | 16 (±2) | >100 | 13 (±1) | >100 | 10 (±1) | >100 |
| **Magnetic ground state**[5] | Canted | | Conical | | Conical | |
| **$T_C$ (K)** | 68(2) [b] | 68 – 75 [60],[16],[6] | 92(2) [b] | 92– 97 [14],[6] | 40(2) [b] | 40 – 51 [7],[6], [61], [14] |
| **$T_s$ (K)**[b] | 15 | 31 [6], [16] | 24 | 24 [14], 27 [6] | 10 | 18 [6], 19.5 [14] |
| **$T_l$ (K)**[b] | | | 5 | 13, 15 [6] | | 14 [14], 15 [6], 17.5 [7] |
| **$T_f$(K)**[b] | 68 | 68 [60] | 88 | 88 [60] | 20 | 20 |
| **Core fraction (%)**[d] | 19 (±2) | 100 | 36 (±2) | 100 | 26 (±2) | 100 |
| **Dead layer $t$ (nm)** | 3.37 (±0.44) | | 1.88 (±0.23) | | 1.81 (±0.26) | |
| **$M_s$ ($\mu_B$/f.u.)**[c] | 0.05 – 0.06 | 0.2 – 0.3[6], 0.3[e][16] | 0.05 – 0.1 | 0.15 – 0.25 [6], [15], 0.33[e][14] | 0.26 – 0.31 | 1 – 1.2 [5], [6], 1.2[e][14], [62] |
| **$\mu_{eff}$ ($\mu_B$/f.u.)** Experiment | 5.88(6) | 10.08 [60] | 6.10(2) | 9.52 [60] | 7.43(7) | 8.06 [61], 8.25[7] |
| Theory | 6.164 | | 6.708 | | 8.062 | |
| **$\theta_W$ ( K)** | - 397(7) | - 536 [60] | - 375(3) | -600[15], -638[60] | - 310 (6) | - 400 [61], -411[7] |
| **$\|\theta_W\|/T_C$** | 5.84(22) | 7.86 [60] | 4.08(5) | 6[15], 6.78[60] | 7.76(29) | 8.7 [61], 10[7] |

a. previously reported results for bulk poly- and single-crystalline samples.
b. as indicated by the derivative $dM(T)/dT$ and ZFC $M(T)$ data measured at $H$ = 100 Oe.
c. reported approximate values observed at low temperatures (2 - 5 K) using bulk samples.
d. results of fitting data of $M(H)$ to Eq. 10
e. results based on neutron diffraction measurements [16].

hyperbolic low-$T$ asymptote with $\xi$ and $\theta'$ are constants. The observed behavior of the magnetic susceptibility of FIM $ACr_2O_4$ NCs above $T_C$ fits well to Eq. (8) as shown by solid lines in insets of Figs. 5(d-f) with $C$ = 4.33(4), 4.65(1) and 6.90(7) emu K /mol, $\theta_W$ = -397(7), -375(2) and -310(6) K, $\xi$ = 1000(41), 604(8) and 823(27) mol K/emu, and $\theta'$ = 54.5(6), 87.3(1) and 25.6(7) K. As the magnetic $Cr^{3+}$ cations are arranged in a pyrochlore sublattice with an inherent tendency of antiferromagnetic Cr-Cr interactions, $ACr_2O_4$ are geometrically spin-frustrated even if the A site is occupied by magnetic ions[14]. The frustration degree is indicated by a ratio $|\theta_W|/T_C > 1$. As shown in Table 3, a frustration factor, $f = |\theta_W|/T_C \approx 5.85$, 4.1 and 7.76 are observed for Ni, Co, and Mn spinel chromite NCs, respectively. The effective paramagnetic moment $\mu_{eff}$ is experimentally estimated as, $\mu_{eff} = \sqrt{3\, k_B C/N_A}$, where $k_B$ and $N_A$ are the Boltzmann's constant and Avogadro's number, respectively. The expected values of $\mu_{eff}$ can be theoretically calculated as,

$$\mu_{eff} = \sqrt{(\mu_{eff})_A^2 + 2(\mu_{eff})_B^2} = g\sqrt{S_A(S_A+1) + 2S_B(S_B+1)}, \tag{9}$$

where $g$ is the Landé g-factor, assumed as 2, and $S_A$ and $S_B$ are the spin quantum number of ions located in the A and B sites, assumed as 1, 3/2, 5/2 and 3/2 for $Ni^{2+}$, $Co^{2+}$, $Mn^{2+}$ and $Cr^{3+}$, respectively. The experimentally determined magnetic parameters of $ACr_2O_4$ ($A$ = Ni, Co, and Mn) NCs are presented with their average particle size, and in a comparison to correponding results reported using bulk poly- and single crystalline samples, in Table 3.

The isothermal magnetic hysteresis loops of $ACr_2O_4$ NCs were measured at different temperatures down to 2 K after a ZFC process and are shown for selected temperatures in Figs. 6(a) – (c) for $A$ = Ni, Co, and Mn, respectively. In agreement with bulk results [15], [52], [57], the magnetization of the three compounds NCs does not saturate with magnetic fields up to 50 and 70 kOe, with the magnetization of Ni and Co chromites are much lower than that of Mn chromite as the case of bulk counterparts. However, the magnetization values are significantly lower than that of bulk sample as shown in Figs. 6(d) – (f). This decrease in $M(H)$ is attributed to an expected significant paramagnetic-like contribution from the surface uncompensated spins in a core-shell FIM particles.

To approximately estimate the contribution fraction of the bulk core magnetization $M_{Bulk}(H)$ to the total net magnetization of the core-shell nanoparticle $M_{Nano}(H)$, we assume that the field dependence of $M_{Nano}(H)$ is given by,

$$M_{Nano}(H) = f\, M_{Bulk}(H) + (1-f)\chi\, H, \tag{10}$$

where $f$ measures the approximate fraction of the core magnetization and $\chi$ is a susceptibility of the surface magnetically disordered (dead) layer. Fitting experimental data $M(H)$ to Eq. (10) at different $T$, where the magnetic state is most stable, i.e. $T_s < T < T_C$, employing the bulk data reported by Mufti et al, Ali and Singh, and Bhowmik et al [6], [62], [63], indicates core contribution fraction with values of about 0.20, 0.35 and 0.25 for Ni, Co and Mn chromite nanoparticles, respectively. As shown in Figs. 6(d) – (f), the estimated $f$ is almost constant with the temperature change.

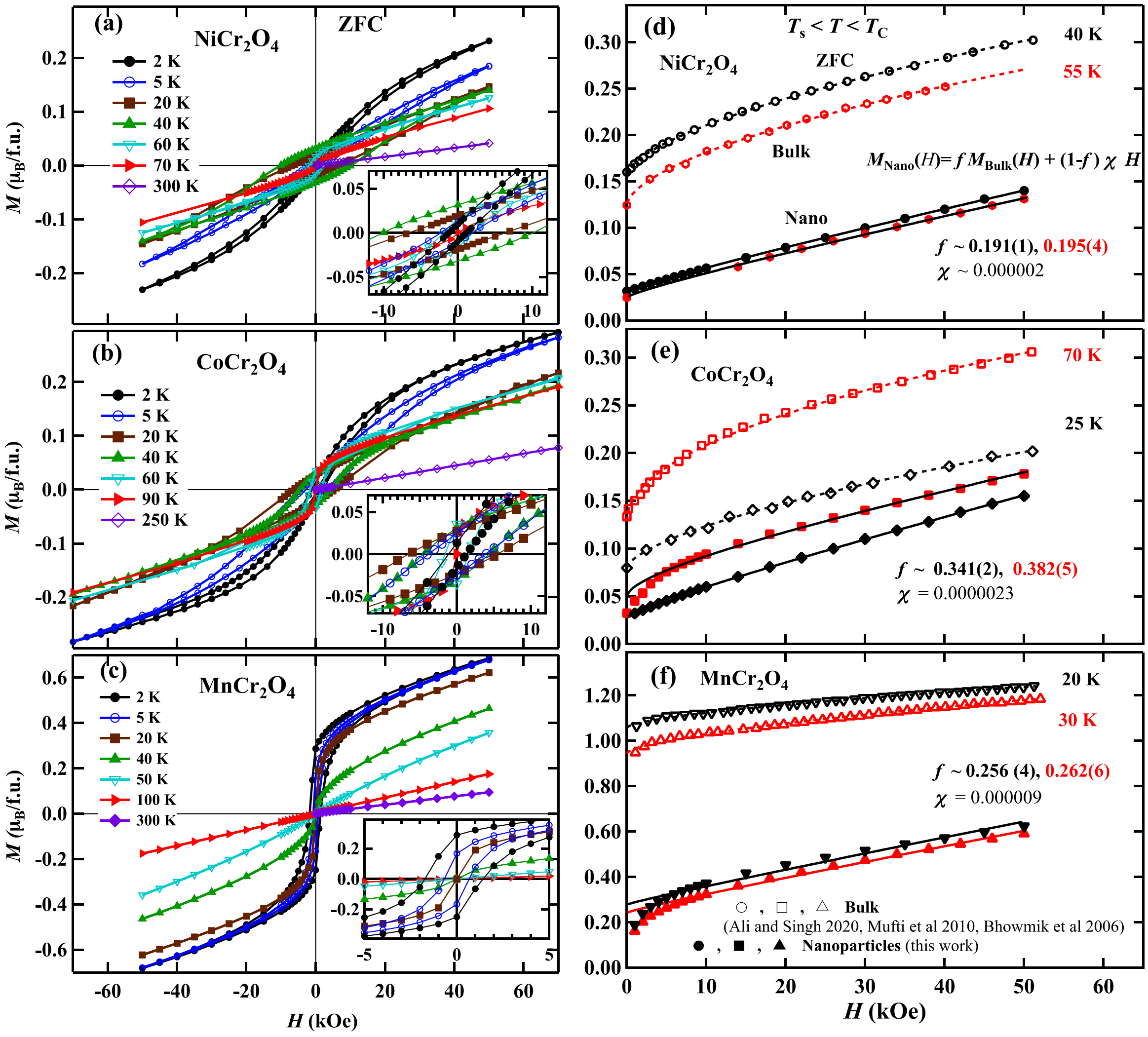

**Figure 6.** The ZFC isothermal magnetic hysteresis loops, (a) – (c), and initial magnetization curves $M(H)$, (d) – (f), for $NiCr_2O_4$, $CoCr_2O_4$, $MnCr_2O_4$ NCs, respectively, shown at selected temperatures. The inset figures of (a) – (c) are magnifications for corresponding main panels at low fields with solid lines for guidance. Bulk $M(H)$ results[6], [62], [63] are shown for comparison in (d) – (f) with dashed lines for guidance while solid lines in (d) – (f) are for best fits to Eq. (10) in the text.

One further finite-size effective property is the magnetically dead layer at the particles surface discussed in a proposed core-shell model, with a particle-size independent thickness that can be subsequently approximated for these chromite NCs[64]–[66]. As the magnetization is proportional to the corresponding volume, we can simply assume that $f = \frac{M_{\text{Bulk}}(0)}{M_{\text{Nano}}(0)} = (\frac{D_{\text{core}}}{D})^3$, where $D_{\text{core}}$ and $D$ are sizes of the core and a total particle, respectively. Hence, the dead layer thickness, $t = (D - D_{\text{core}})/2$ can generally be estimated as:

$$t = \frac{D}{2}(1 - f^{1/3})\ . \qquad (11)$$

Within the accuracy limits of the determination of particle/crystallite size $D$ as well as volume fraction $f$, average values of $t$ are approximately estimated as 3.4(4), 1.9(2) and 1.8(2) nm for the Ni, Co, and Mn chromite nanoparticles of average particle size 15, 13 and 10 nm, respectively. Values of $t$ (~ 1 – 3 nm) close to values approximated here in reasonable estimates have been reported for manganite and ferrite nanoparticles[66]–[68]. Referring to the lattice constant values, Table 1, a shell layer of ~4$a$ in thickness for the Ni and ~2$a$ for the Co and Mn chromite NCs is magnetically dead at the surface of these nanoparticles. For example, by assuming the spherical shape of these nanoparticles, roughly about 2500 among 3000 unit cells per a particle in the $NiCr_2O_4$, 1300 among 2000 in the $CoCr_2O_4$, and 750 among 1000 in the $MnCr_2O_4$ NCs have no net magnetic moment, which reflects the highly reduced spontaneous magnetization in these nanoparticles.

Characteristics of the magnetic hysteresis loop such as the squareness shape and ratio and the coercive field, and their evolution with temperature in an interrelation to the freezing features can provide information about the driving interactions in these spinel FIM nanoparticles. The strength of a spin-glass-like state observed below a freezing temperature, $T_f$, can be indicated by the difference $M_{FC}$-$M_{ZFC}$ at low fields. The squareness ratio is estimated as SQR = $M_r$ / $M_s$, where $M_r$ is the observed remanent magnetization and $M_s$ is the high-field 'saturation' magnetization. $M_s$ is approached here by extrapolating linear fits of the high-field magnetization $M(H)$ data to $1/H^2 = 0$ in $M(H)$ vs $1/H^2$ plots, as shown in the inset of Fig. 6(d) for the data at 2 K. Figure 7 shows the temperatures dependence of the $M_{FC}$-$M_{ZFC}$ (at 100 Oe), SQR and $H_c(T)$ of $ACr_2O_4$ ($A$ = Ni, Co, and Mn) NCs accompanied with reproduced results previously reported using bulk single and polycrystalline samples [6], [14], [37], [52], [56], [57], [60], [62], [69]. Below $T_f$, the difference $M_{FC}$-$M_{ZFC}$ qualitatively reproduces those of bulk polycrystalline samples showing anomalies corresponding to each compound magnetic transitions, as well as for $M_{FC}$-$M_{ZFC}$ at 10 kOe shown in Figs. 5(d) and (f). Small variation of the $M_{FC}$-$M_{ZFC}$ of Mn NCs from that of the bulk behavior exhibiting anomaly features around the bulk $T_s$, Figs.7(g), can be attributed to the suppression of its spiral spin order by finite-size effects.

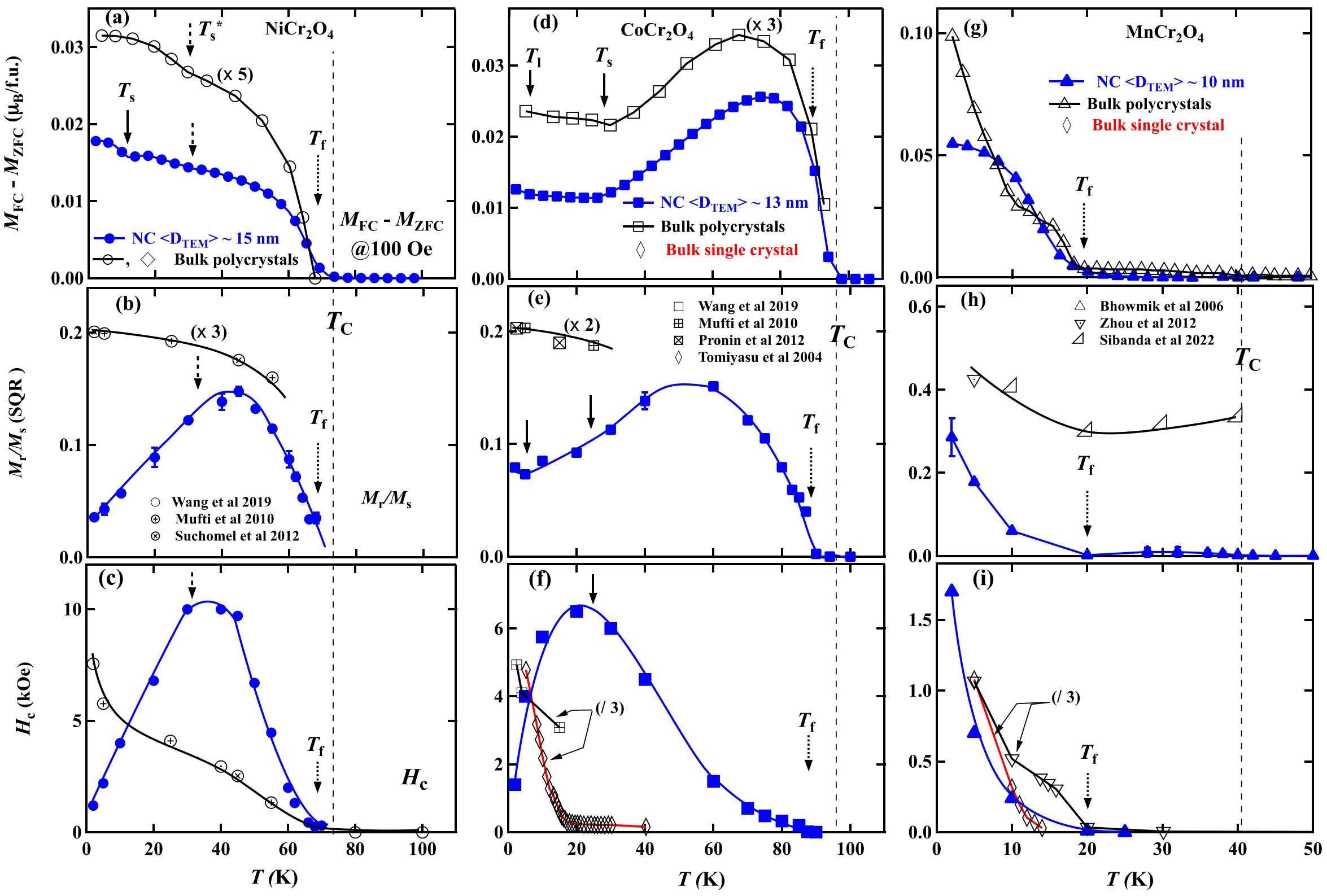

**Figure 7.** The temperature dependences of the FC magnetization splitting from ZFC magnetization, $M_{FC}$-$M_{ZFC}$, measured at 100 Oe, the estimated squareness ratio (SQR = $M_r$ / $M_s$), and observed coercivity, $H_c$, for **(a - c)** $NiCr_2O_4$, **(d – f)** $CoCr_2O_4$ and **(g– i)** $MnCr_2O_4$ nanocrystals (closed symbols), presented in comparison to those of bulk samples (open symbols), reproduced from Refs. [6], [52], [57], [60], [62], [69] for polycrystalline samples and from Ref. [14] for single crystals. Solid lines are for a purpose of eye guidance.

Unlike the bulk results, unusual nonmonotonic behaviors are observed in the SQR($T$) and $H_c(T)$ of Ni and Co chromite NCs below $T_f$. Broad peak with large coercivity values is around the onset of the transitions to noncollinear, canted AFM and spiral FIM, in Ni and Co chromite NCs, respectively. The SQR increases up to 0.15 before it decreases with decreasing the temperature below $T_s$ for Ni and Co chromite NCs, while it continuously increases to ~ 0.3 at the lowest temperature for the Mn chromite NCs. Similar behavior is observed for $H_c(T)$ with maximum values of about 10 and 6 kOe around $T_s$ for Ni and Co chromite NCs, respectively, and continuous increase up to ~ 2 kOe at 2 K, for the Mn chromite NCs.

## 4. Discussion

In this section we discuss the interrelations and effects of the geometric frustration, the electronic structure, and the finite size on the magnetic properties of the non-collinear FIM Ni, Co, and Mn chromite nanocrystals in a view to results reported for bulk samples. The spinel chromites crystalize in a normal spinel structure with high stabilization energy when $Cr^{3+}$ fully occupy the octahedral *B* sites. As indicated by of results XRD analysis in section 3.1, the normal spinel structure is not affected by decreasing the particle size of these materials down to 10 nm, in agreement with previously reported results[53], [62], [70]. In this structure, a pyrochlore sublattice purely formed by $Cr^{3+}$ is associated with a geometrical magnetic frustration usually yielding to a nontrivial magnetic ground state in a FIM spinel chromites. We found that the synthesized NCs exhibit a wide optical energy gap for $CoCr_2O_4$ NCs (3.1 eV) and gradually decreases for $MnCr_2O_4$ (2.9 eV) and $NiCr_2O_4$ NCs (2.87 eV). In a previously reported study of the magnetic properties of spinel chromites in relation to their electronic structures, it was found that the spinel chromite with a relatively large optical energy gap can undergo a first-order phase transition into a spiral spin ground state[5]. In this sense, the spiral spin-order transition is significantly observed for $CoCr_2O_4$ NCs, with a relatively wider energy gap, rather than for the $MnCr_2O_4$ NCs. On the other hand, $NiCr_2O_4$ NCs exhibit their bulk canted AFM ground state as described above in section 3.3. However, the bulk magnetic ground state of a spinel chromite was basically attributed to the degree of geometrical spin frustration in the $Cr^{3+}$ pyrochlore sublattice that is determined by the $A^{2+}$ ion and its magnetic moment [13].

In a theory proposed by Lyons, Kaplan, Dwight, and Menyuk (LKDM)[13], the magnetic ground state of spinels with spiral or conical structures can be well described by a parameter *u* is given by $u = \frac{4\,J_{BB}S_B}{3\,J_{AB}S_A}$. In this theory, $J_{BB}$ and $J_{AB}$ are the only nearest-neighbor antiferromagnetic interactions among spins $S_B$ and $S_A$ at the A and B sites, assuming a weak A-A magnetic interaction. As the spin order is modified by the inherent geometrical magnetic frustration of the spinel B site, the parameter *u* takes possible values ranges from 0 to infinity corresponding to spin configurations between the Neél type FIM state and a state of highly frustrated spins, respectively[6]. In a spinel with $u < 8/9$, the ground state is a long-range-ordered (LRO) Neél-type FIM state. With larger $u$, a LRO spiral state evolves and further it becomes locally unstable yielding to a SRO spiral state in the region where $u > 1.298$. The values of *u* corresponding ground states of $CoCr_2O_4$ and $MnCr_2O_4$ are 2 and 1.5, respectively, while that of the canted $NiCr_2O_4$ is expected to be relatively larger. Bulk $CoCr_2O_4$ and $MnCr_2O_4$ exhibit a conical magnetic structure that is stabilized by a SRO incommensurate spiral component develops below $T_s$ and becomes commensurate to the lattice at a 'lock-in' transition temperature $T_l$. Based on neutron scattering measurements using single-crystal samples of $CoCr_2O_4$ and $MnCr_2O_4$ [14], [54], the spiral component emerges below $T_s$ = 25 and 18 K, with coherence length reaches 3.5 and 10 nm, in propagation vectors $Q$ = (0.62, 0.62, 0) and (0.59, 0.59, 0), respectively. On the other hand, in $NiCr_2O_4$, a transverse antiferromagnetic component appears below $T_s$ with a propagation vector $Q$ = (0, 0, 1) and coplanar (but non-collinear) spin configuration has been revealed by neutron scattering [16]. The variation of $NiCr_2O_4$ from $CoCr_2O_4$ and $MnCr_2O_4$ in the magnetic ground state may be attributed to the large tetragonal distortion occurs by a Jahn−Teller transition, observed at 310 K in bulk samples [16]. Among the divalent cations $A^{2+}$ in the three compounds studied here, only $Ni^{2+}$, with the spin configuration shown in Fig. 1(a), exhibits an orbital contribution to its magnetic moment with a net ionic moment of 3 $\mu_B$. Considering the magnitude of magnetic moments of $Ni^{2+}$, $Co^{2+}$ and $Cr^{3+}$ at its theoretical value of 3 $\mu_B$ and that of $Mn^{2+}$ at 5 $\mu_B$, spin structure models represented by Tomiyasu et al based on neutron scattering results [14], [16], almost reproduce the experimental bulk magnetic moments observed for Ni, Co and Mn spinel chromites.

The nanoparticles form of these FIM spinel chromites interestingly enables to study their magnetocrystalline anisotropy in a single domain and its evolution with temperature rather than in bulk-form samples. Very small magnetic particles must be single-domain and below a certain critical size they do not benefit energetically from the wall formation. Assuming almost spherical particles, we can approximately estimate the single-domain maximum size as $2R_{sd} = \frac{18\sqrt{A\,K_u}}{\mu_0 M_S^2}$ [71], where $A = \frac{3\,K_B T_C}{z\,a}$ and $K_u = \frac{H_c\,M_S}{0.96}$, are the exchange stiffness and the anisotropy constant, respectively, assuming the Stoner–Wohlfarth model for randomly oriented identical particles[72], [73]. $M_s$ is the saturation magnetization, $K_B$ is the Boltzmann constant, and z is the effective number of the nearest neighbors, with all quantities are in SI units. Applying these criteria to $MnCr_2O_4$ NCs as the most stable FIM spinel chromite in this study, indicates a critical size of its single domain in the order of 0.1 μm. This approximate estimation indicates that our spinel chromite nanoparticles are single-domain magnetic particles that free from domain walls.

The anomalous magnetic behaviors observed in the temperature dependence of the coercive field, $H_c(T)$, and the hysteresis-loop squareness, Fig. 7, are indicative for the magnetocrystalline anisotropy of these FIM spinel chromite NCs. As the nanoparticles qualitatively reproduce the freezing behavior of the bulk samples, we can exclude the effect of the surface spin disorder on the magnetic anisotropy. For $NiCr_2O_4$ and $CoCr_2O_4$ single-domain NCs, with decreasing the temperature below $T_C$ a net moment of a collinear FIM order increases in a preferred direction simply resulting in an increased anisotropy and hence the coercive field required for a coherent magnetization reversal, Figs. 7(b) and (e). At and below $T_s$ the emergence of a noncollinear magnetic order, canted AFM and spiral FIM in $NiCr_2O_4$ and $CoCr_2O_4$, respectively, reduces the magnetic anisotropy, and hence the coercive field, that gradually decreases with a possible increase in the canting angle with decreasing the temperature. The implied softness of the magnetic structure below $T_s$ influences the saturation and remanent magnetizations resulting in similar temperature dependencies of the hysteresis-loop squareness shape, Figs. 7(b) and (e). In addition, the unusual behavior of the squareness ratio and the coercive field at low temperatures could be related to some exchange bias of the surface moments with the core moments. On the other hand, the $MnCr_2O_4$ fine NCs with most stable spin structure exhibits a monotonic magnetic behavior almost similar to its corresponding bulk results with addition effects of the surface disorder that suppresses the spiral transition, Figs. 7(g -i).

Finally, the nanocrystalline FIM spinel chromite of current study exhibits crystallite size effects on the complex non-collinear spin order. The approximate fraction of the core magnetization with estimated values of about 20, 36 and 25 %, for Ni, Co, and Mn chromites, respectively, can explain the observed suppression of the spiral order in these chromite nanoparticles, with different extends in the different chromites. For example, a clear crystallite size dependent transition ($T_s$) to a canted AFM state is observed in $NiCr_2O_4$ and the coexistence of particles with different sizes may result in additional anisotropic exchanges. For $CoCr_2O_4$ NCs, almost bulk transition to the incommensurate spiral spin order was observed, however, its lock-in magnetic transition ($T_l$) exhibits a crystallite size effect. On the other hand, the magnetic transitions to incommensurate and commensurate spiral orders are suppressed in $MnCr_2O_4$ NCs with relatively smaller particle sizes. Instead, a coexisting spin-glass-like state is observed with a FIM state below $T_C$. The presented results spotlight on the interrelation between the crystal and electronic structures in finite sizes and the unusual magnetic behavior of FIM spinel chromites. A study of the particle-size dependence of these magnetic features and related phenomena such as the exchange bias effect in these materials are stimulated by the currently presented results.

## 5. Conclusion

Single-phase FIM $ACr_2O_4$ ($A$= Ni, Co, and Mn) NCs were synthesized by a one-pot microwave-combustion method. Room-temperature cubic $Fd\bar{3}m$ normal spinel-structured phase was indicated for the three compounds NCs by the XRD and FT-IR results. Compared to results using bulk samples, the UV-Vis absorption demonstrated additional 3d-3d transitions in the Co and Ni chromite NCs spectra, unlike the Mn chromite NCs, with an estimated wide optical band gap of ~3 eV. Finite size effects on the low-temperature noncollinear FIM transitions and intriguing single-domain anisotropic features have been revealed in this study. $NiCr_2O_4$ NCs of average particle size 15 nm exhibit a canted AFM transition around $T_s$ = 15 K which is much lower than that in single crystals, while $CoCr_2O_4$ NCs of 13 nm size show almost the bulk transition to an incommensurate spiral spin order at $T_s$ = 24 K, however, its lock-in magnetic transition appears at $T_l$ = 5 K is shifted below the bulk transition (15 K). On the other hand, the incommensurate spiral FIM transition appears at 10 K, shifted below the bulk transition (20 K), while its lock-in is completely suppressed in $MnCr_2O_4$ NCs of average particles size 10 nm. The observed anomalous behaviors of the coercive field $H_c(T)$ and hysteresis-loop squareness ratio SQR($T$), with broad peaks around $T_s$ of $NiCr_2O_4$ and $CoCr_2O_4$ NCs, indicate a single-domain magnetic behavior and elucidate the magnetotcrystalline anisotropy of these materials.

## Acknowledgment

The authors would like to thank N. Sasaki and K. Kazumi from the Department of Materials Science and Engineering, Kyoto University, for their help with the TEM analyses. M.A.K sincerely thanks the Japan Society for the Promotion of Science (JSPS) for Postdoctoral Fellowship for Research in Japan (Kyoto University). A.M.N would like to thank the Ministry of Higher Education and Scientific Research (Egypt) for a scholarship support to Kyoto University (Japan).